\documentclass[a4paper,fleqn,usenatbib]{mnras}

\usepackage{graphics}
\usepackage{graphicx}
\usepackage{psfig}
\input{epsf}
\usepackage{amsmath}
\usepackage{amssymb}
\usepackage{mathtools}
\usepackage{url}
\usepackage[T1]{fontenc}
\usepackage{ae,aecompl}
\usepackage{widetext}
\usepackage{hyperref}
\usepackage{cleveref}

\newcommand{\be}{\begin{equation}}
\newcommand{\ee}{\end{equation}}

\newcommand{\bea}{\begin{eqnarray}}
\newcommand{\eea}{\end{eqnarray}}
\newcommand{\bean}{\begin{eqnarray*}}
\newcommand{\eean}{\end{eqnarray*}}

\newcommand{\lcdm}{$\Lambda$\text{CDM}\,}
\newcommand{\planck}{\emph{Planck} }

\newcommand{\GN}{G_N}
\newcommand{\cmb}{\text{{CMB}}\,}
\newcommand{\isw}{\text{ISW}\,}
\newcommand{\md}{\mathrm{d}}
\newcommand{\mathH}{\mathcal{H}}

\title[ISW effect in the BVDE cosmology]{The Integrated Sachs-Wolfe Effect in the Bulk Viscous Dark Energy Model}

\author[B. Mostaghel, H. Moshafi and S. M. S. Movahed]{B. Mostaghel$^{1}$, H. Moshafi $^2$ \& S. M. S. Movahed$^{1,2}$\thanks{email:\href{mailto: b{\_} m.s.movahed@ipm.ir} {m.s.movahed@ipm.ir}}
\\
\\
$^{1}$ Department of Physics, Shahid Beheshti University, Velenjak, Tehran 19839, Iran\\
$^{2}$ Ibn-Sina Laboratory, Shahid Beheshti University, Velenjak, Tehran 19839, Iran
}

\begin{document}
\label{firstpage}
\pagerange{\pageref{firstpage}--\pageref{lastpage}}
\maketitle

\begin{abstract}
 We examine  linear perturbation theory to evaluate the contribution of viscosity coefficient in the growing of dark matter perturbations in the context of the bulk viscous dark energy model  inspired by thermodynamical dissipative phenomena proposed by  \cite{Mostaghel:2016lcd}. As the cosmological implementations, we investigate the {\it Integrated Sachs-Wolfe} (ISW) auto-power spectrum, the ISW-galaxy cross-power spectrum and derive limits on $f\sigma_8$. The dimensionless bulk viscosity coefficient ($\gamma$) in the absence of interaction between dark sectors, modifies the Hubble parameter and the growth function, while  the Poisson equation remains unchanged. Increasing $\gamma$  reduces  the dark matter growing mode at the early epoch while a considerable enhancement will be achieved at the late time. This behavior imposes non-monotonic variation in the time evolution of gravitational potential generating a unique signature on the CMB photons. The bulk viscous dark energy model leads to almost a decreasing in ISW source function at the late time. Implementation of the Redshift Space Distortion (RSD) observations based on "Gold-2017" catalogue,  shows $\Omega^0_{\rm m} =0.303^{+0.044}_{-0.038}$, $\gamma=0.033^{+0.098}_{-0.033}$ and $\sigma_8= 0.769^{+0.080}_{-0.089}$ at $1\sigma$ level of confidence. Finally, tension in the $\sigma_8$ is alleviated in our viscous dark energy model.
\end{abstract}


\begin{keywords}
methods: analytical--methods: data analysis -- cosmic background radiation -- dark energy -- large-scale structure of Universe.
\end{keywords}


\section{INTRODUCTION}\label{intro}

The agent of the late time accelerating expansion of the Universe confirmed by many observational data sets ranging from background evolution to perturbations dynamics is  mysterious not only for theoretical cosmology but also in observations \citep{Riess:1998cb,Perlmutter:1998np}.

The standard cosmological model  (\lcdm) containing six free parameters is a trivial prescription to explain dynamics of our Universe. This scenario has been confirmed by various observations such as Supernova Type Ia (SNIa), Cosmic Microwave Background radiation (CMB), Baryonic Acoustic Oscillations (BAO) \citep{Ade:2015xua,Ade2016}, the ISW effect \citep{boughn2004correlation,boughn2005cross,boughn2005detection,ade2016planck} and weak lensing \citep{Peel:2016jub,heymans2013cfhtlens,Lewis:2006fu,contaldi2003joint}. Despite the outstanding consistencies between \lcdm and the observational cosmic data sets, the cosmological constant has some fundamental problems and concordance model has tensions remained unresolved \citep{Ade:2015xua,Riess2016,Bernal2016}. Recent observations indicated a deficiency in amplitude of ISW cross-correlation with astronomical objects based on concordance model \citep{Granett2008,Kovacs2013,Flender2013,ferraro2015wise}.  Taking into account the late-Universe measurements of the dark matter growth rate which is proportional to the $\sigma_8$ implied an orientation to low value with respect to that of computed by CMB in the context of \lcdm \citep{ade2016planck,Heymans:2012gg,erben2013cfhtlens}. Subsequently, there is some room for alternative scenarios mainly classified into the following categories: Dynamical dark energy including the field theory orientation, phenomenological dark fluids, modification of the general relativity and  thermodynamics point of view \citep[and references therein]{Copeland:2006wr,Amendola:2012ys,Horndeski:1974wa,Konnig:2016idp,Bento:2002ps,Zlatev_1999,Caldwell:1999ew,Amendola:1993uh,Germani:2010gm,Huterer:2017buf,Mostaghel:2016lcd}.  In an interesting approach, recently, N. Khosravi suggested a new proposal to modify the standard cosmology based on the idea of taking ensemble average over the various gravitational models \citep{Khosravi2017,khosravi2017uber,Khosravi2016}.  The $\Lambda(t)$ cosmology, considering a typical form of dark energy and/or interaction between dark sectors in the Universe are  some proposals to reduce such discrepancies \citep{Wang2010a,Velten2015,Kunz_2015,barbosa2017assessing,fan2016integrated,mainini2011isw}.


Another proposal for dark energy is an exotic fluid with some
thermodynamical features such as bulk and/or shear viscosities. 
Meanwhile, finding proper observational measures which can precisely probe the influence of dark energy component and  distinguishing between various scenarios have received extensive attention. Geometrical and topological properties of cosmological random fields \citep{Chenxiaoji:2014mxa,Fang:2017daj} and considering the primary and the secondary probes have been discovered and applied for evaluation and discrimination of dark energy models \citep[and references therein]{Huterer:2017buf}.

Dark energy can affect on the various elements of the Universe. A trivial contribution of dark energy can be realized in the rate of background expansion. Hence all quantities containing the Hubble parameter are affected by dark energy density.  At the background level, due to changing the expansion history of the Universe, the distance to the last scattering surface is modified changing the so-called  acoustic signatures   \citep{Hu1996}. Dark energy perturbations can also alter the lensing potential \citep{Acquaviva2006,Carbone2013} and consequently can modify the lensing $B$-mode \citep{Amendola2014}.  Dark energy modifies the primordial gravitational wave and therefore it changes the amplitude of primordial $B$-mode \citep{Antolini2013,Raveri2015,Amendola2014}.  The matter perturbations growth is also affected by dark energy density \citep{Peebles1984,Barrow1993} causing to have some discrepancies between amplitude of fluctuations computed by late-time observations and CMB map \citep{Durrer1999,Baldi2011,Ade2016}.

Secondary CMB anisotropies are widely experiencing different epoch of the Universe, therefore, we expect that the CMB stochastic field is a proper measure to explore dark energy. A relevant probe of dark energy in the context of CMB observations, is  ISW effect which is a secondary anisotropy \citep{Sachs1967,M.J.1968,Kofman1985, Hu:1993xh,crittenden1996looking,Cooray:2002ee,afshordi2004integrated,Schaefer:2005up,Schaefer2008,ade2016planck}. ISW effect is related to the frequency changes in the CMB photons when they encounter with the time evolving gravitational potential.  Since dark energy is dominating at the low redshift\footnote{It turns out that various dark energy cosmological models can provide different range of redshifts depending on corresponding natures.} (${z\sim 1}$), the primordial \cmb fluctuations alone cannot provide a considerable and precise probe. However, the secondary anisotropies in the \cmb is  more sensitive to the dark energy dynamics \citep{Schaefer2008,Kovacs2013,Huterer:2017buf}. The cosmic variance for almost those multipoles that ISW has signature on the CMB power spectrum washouts the importance of ISW alone to explore dark energy models \citep{Song:2006jk}.

 Practically, the cross-correlation of ISW with the tracers of the large scale structures magnifies the ISW signal and it would be distinguishable from primordial processes \citep{afshordi2004cross,afshordi2004integrated,ho2008correlation,olivares2008integrated,Giannantonio:2008zi, douspis2008optimising,Wang2010a,ferraro2015wise,Lesgourges2013}.

The ISW signal has been computed, in order to examine  the $\Lambda$ contribution in concordance model relying on  rare superstructures identified in the SDSS Luminous Red Galaxy catalogue \citep{nadathur2012integrated,Flender2013}. The ISW effect and its cross-correlation with large scale structures  have been investigated in the context of alternative to \lcdm models such as the extended quintessence model in both the metric and Palatini formalisms \citep{fan2016integrated}, quintessence cosmological model \citep{Wang2010}, particular form of interaction between dark sectors  \citep{Schaefer2008,schafer2008mixed,malte2009implications},  non-ideal fluid dark energy with anisotropic stress component  \citep{majerotto2015combined}, clustering of dark energy \citep{khosravi2016isw} and also other more general cosmological models \citep{scranton2003physical,ferraro2015wise,Velten2015,fosalba2003detection,dent2009new,sapone2009fingerprinting,padmanabhan2005correlating,dent2009new}.

More
recently, inspired by thermodynamical dissipative phenomena and
taking into account the isotropy of the Universe at the background
level, we proposed a bulk viscous dark energy (BVDE) model
\citep{Mostaghel:2016lcd}. In this model, the old cosmological
objects can be accommodated and the tension in the Hubble parameter
was alleviated \citep{Mostaghel:2016lcd}. Relying on dissipative process in a realistic fluid, Israel et al. proposed a causal dissipative theory for assessing the irreversible processes
\citep{Israel_1979}. Accordingly, bulk and shear viscous terms are most relevant parts for a feasible relativistic fluid.  The Ekart's theory including the first-order dissipative relativistic fluid, is acausal and has instabilities \citep{Eckart:1940te,Hiscock:1985zz,Hiscock:1991sp} (see also \citep{landau1987fluid,Jou2009}). There are many approaches to construct causal and stable theory of relativistic viscous fluid for a certain range of relevant quantities and proper conditions \citep{hiscock1983stability,hiscock1988nonlinear,disconzi2014well}. However, examining the problem of causality and stability of relativistic theories is under debate \citep{rezzolla2013relativistic}.  
 
For cosmological implementation, it has been demonstrated that Israel approach converges to the Eckart's theory  \citep{Hiscock:1991sp}. Since, the collision time scale in the transport equation of our proposed fluid is zero, consequently, our bulk viscous model is necessarily acausal and unstable \citep{Maartens:1996vi}. For making  a causal bulk viscous dark energy, in principle, one should take into account full Israel-Stewart transport equation and keeping the collision time scale \citep{Israel_1979}.  On the other hand, the functional form of viscosity in our model leads to crossing phantom divide barrier which has been observationally confirmed \citep{Ade2016}. In order to resolve mentioned instability and regularize underlying model, in some cases, one can take into account interaction between dark components of the Universe \citep{Amendola:2012ys}. The effective dark energy in a suitable interacting model can possibly have phantom crossing without divergences and it is fundamentally related to the collision time scale in the corresponding transport equation. It is worth noting that in a simple dark energy model with constant equation of state accompanying a typical interaction term leads to another instability \citep{valiviita2008large}. One approach to reduce the effect of instability may possibly be assumed a proper rate of interaction between viscous dark energy and dark matter. Another way for regularization the instability is suppressing any initial perturbations of dark energy by taking proper initial conditions.

 In this paper, we are interested in discussing the consequence of BVDE model incorporating the background expansion on the large scale structures. To this end, we will take into account the linear perturbation of dark matter in the presence of our viscous dark energy model to study its contributions on the rate of structure formations and ISW effect \footnote{Since there are various types of bulk viscosity models for dark energy cosmologies, some authors have tried to evaluate linear structure formation process \citep[and references therein]{barbosa2017assessing}.}.

  We will focus on the ISW auto-power spectrum and  cross-correlation between CMB fluctuations mainly considered by ISW part with large scale structure to explore the contribution of bulk viscosity in the dark energy budget. To make our discussion more complete,  we will use the Redshift Space Distortion (RSD) data set to constrain model free parameters.  The contribution of viscosity coefficient represents a non-monotonic behavior of ISW auto- and cross-power spectrums leading to low clustering compared to \lcdm model and also it can reduce the tension in $\sigma_8$. It is worth reviewing that the BVDE cosmological model could alleviate $H_0$ tension \citep{Mostaghel:2016lcd}.

 The layout of the rest of this paper is as follows:  In section \ref{sec:BVDE}, for the sake of clarity, we give a brief introduction of our bulk viscous dark energy (BVDE) model. Background dynamics of the Universe will be explained in this section. Section \ref{sec:structure} is devoted to the linear perturbation of the dark matter when bulk viscous dark energy is considered. Growth function, growth rate  and bias independent parameter, namely $f\sigma_8$ for BVDE model are derived in section \ref{sec:structure}. Angular auto-power spectrum of ISW and ISW-galaxy power spectrum in the flat sky approximation for BVDE model will  be derived in section \ref{sec:ISW}. We will check the observational consistency using RSD data set in section \ref{sec:observation}.  Summary and concluding remarks are given in section \ref{sec:summary}.

\section{Cosmic evolution in the bulk viscous dark energy framework}\label{sec:BVDE}
\begin{figure}
\centering
\includegraphics[width=1.0\columnwidth]{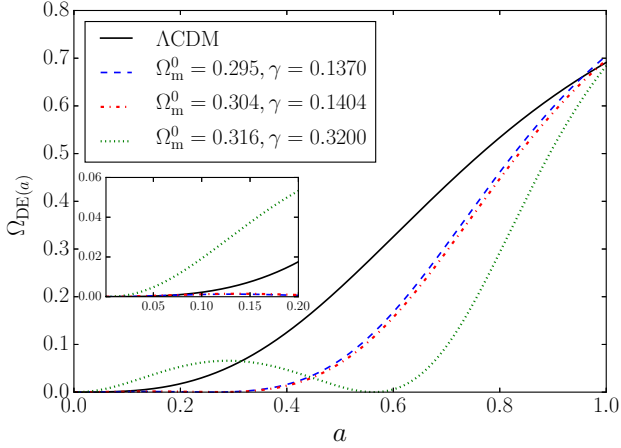}
\caption{The energy density of dark energy in the \lcdm and BVDE models as a function of scale factor. The values of the free parameters selected in this plot are compatible with background tests \citep{Mostaghel:2016lcd}.}
\label{fig:Om_DE}
\end{figure}

A proposal for dark energy model inspired by dissipative fluid dynamics is considering bulk and shear viscous terms for the energy-momentum tensor. Recently, we consider an exotic dissipative fluid playing a responsible for the late time acceleration \citep{Mostaghel:2016lcd}. The pressure for dark energy (DE) component in the BVDE model is defined by:
\bea
p_{\rm DE}=-\rho_{\rm DE}-\zeta \Theta(t),
\eea
where $\zeta = \zeta(\rho_{\rm DE})$ and $\Theta(t)=\nabla^{\mu}u_{\mu}$ are viscosity and expansion scalar, respectively. In the FLRW space-time, expansion scalar is $\Theta(t)=3H(t)=3\dot{a}/a$. In addition, we have an assumption  \citep{Mostaghel:2016lcd}:
\begin{eqnarray}\label{ansatz}
\zeta(\rho_{\rm DE},H)=\xi\frac{\sqrt{\rho_{\rm DE}}}{H},
\end{eqnarray}
where $\xi $ is positive constant with \textit{mass square} dimension. Other functional form for $\zeta$ can be found in \citep[and references therein]{Li:2009mf,Hiscock:1991sp,barbosa2017assessing}. By solving the continuity equation, the evolution of dark energy component becomes:
\bea
\rho_{\rm DE}\left(a\right) = \rho_{\rm DE}^{0}
\left(1 +\frac{9 \xi}{2\sqrt{\rho^{0}_{\rm DE}}}\ln{a}            \right)^{2}.
\eea
The superscript "0" represents the value of DE component at present time. Using the flat FLRW metric in the Einstein's fields equations, we find the Friedmann equations as:
\begin{subequations}
\begin{align}
&H^2= \frac{8 \rm{\pi}\GN}{3}\rho_{\rm tot}, \label{eq:friedmann_1}\\
&\frac{\ddot{a}}{a}=-\frac{4\rm{\pi} \GN}{3}\left ( \rho_{\rm tot}+3 p_{\rm tot}\right )\label{eq:friedmann_2},
\end{align}
\end{subequations}
where $\rho_{{\rm tot}}=\rho_{\rm r}+\rho_{\rm b}+\rho_{\rm m}+\rho_{\rm DE}$  and $p_{{\rm tot}}=p_{\rm r}+p_{\rm b}+p_{\rm m}+p_{\rm DE}$. The $G_N$ is Newton's constant. The indices $"{\rm r}"$, $"{\rm b}"$ and $"{\rm m}"$ correspond to radiation, baryonic matter and cold dark matter (CDM), respectively. Finally,  the Hubble parameter without any interaction between dark sectors reads:
\begin{align}\label{eq:Hubble_1}
\begin{split}
H^2=&H_0^2\left[\Omega_{\rm r}^{0}a^{-4}+\Omega_{\rm m}^0 a^{-3}+\Omega_{\rm DE}^0\left(1 +\frac{9 \gamma}{2\sqrt{\Omega^{0}_{\rm DE}}}\ln{a}\right)^{2}\right]\\
&+H_0^2(1-\Omega_{\rm tot}^0)a^{-2},
\end{split}
\end{align}
where $\Omega_{\rm tot}^0=\Omega_{\rm r}^0+\Omega_{\rm
m}^0+\Omega_{\rm DE}^0$ and throughout this paper, we consider flat
Universe ($\Omega_{\rm tot}=1$). In Eq.~(\ref{eq:Hubble_1}), we define:
\bea
\Omega^0_{\rm r}\equiv\dfrac{8\rm{\pi} \GN}{3H_0^2}\rho^0_{\rm r},\quad
\Omega^0_{\rm m}\equiv\dfrac{8\rm{\pi} \GN}{3H_0^2}\rho^0_{\rm m},\quad
\Omega^0_{\rm DE}\equiv\dfrac{8\rm{\pi} \GN}{3H_0^2}\rho^0_{\rm DE},
\eea
and the dimensionless viscosity coefficient, $\gamma$ is:
\begin{eqnarray}\label{gamma}
\gamma\equiv \sqrt{\frac{8\rm{\pi} G_{N}}{3H_{0}^2}}\xi.
\end{eqnarray}
It turns out that for $\gamma=0$, the standard \lcdm model is recovered. Due to viscosity term, we have non-trivial behavior for DE in BVDE model compared to cosmological constant.  In Fig. \ref{fig:Om_DE}, we indicate the evolution of $\Omega_{\rm DE}(a)$ as a function of scale factor. As indicated in the inset plot of this figure, around $a\lesssim 0.06$ the contribution of DE in the BVDE model is more than $\Lambda$. While for the interval $0.06\lesssim a \lesssim 0.9$ the value of $\Omega_{\rm DE}$ is less than \lcdm model. At very early epoch, again the $\Omega_{\rm DE}$ is similar to $\Lambda$. Subsequently,  this kind of behavior has non-trivial impact on the ISW as well as other structure formation phenomena. In the next section, we will study the structure formation in the BVDE framework.

\section{Structure Formation in the BVDE model}\label{sec:structure}
Cosmic fluctuations seeded by inflationary models and/or topological phase transitions at the early Universe, are evolving through the aging of the Universe and finally form the large scale structures.
 In this section we explain the main part of structure formation and associated physical parameters such as dark matter power spectrum in the presence of BVDE model.

In order to study the evolution of the large scale structures in the Universe filled by cold dark matter (CDM) and viscous dark energy, we consider Einstein equations for small inhomogeneities as:
\begin{equation}
\delta G^{\mu}_{\nu}=8\rm{\pi} G_{N}\delta T^{\mu}_{\nu}.
\end{equation}
The perturbed metric in the Newtonian gauge  for the homogeneous and isotropic Universe is therefore written by (\cite{Bardeen1980,Kodama1984,Mukhanov1992}):
\begin{equation}
\md s^2 =a^2(\eta)\left[-(1+2\Psi)\md \eta^2+(1+2\Phi)\delta_{ij}\md x^i \md x^j\right],
\end{equation}
where $\eta$, $\Psi$ and $\Phi$ are conformal time and metric perturbations, respectively. In absence of anisotropic stress, we have $\Psi=-\Phi$.
Applying the above line element in the Einstein's equations leads to the perturbed gravitational field equations \citep{Bardeen1980,Kodama1984,Mukhanov1992}:
\begin{subequations}
\begin{align}
\delta G^0_0&=2a^{-2}[3\mathH(\mathH \Psi -\Phi')+\Box \Phi] ,\\
\delta G^0_i &=2a^{-2}(\Phi' -\mathH \Psi)_{|i},\\
\delta G^i_j&=2a^{-2}[(\mathH^2 +2 \mathH ')\Psi  + \mathH \Psi '-\Phi '' -2\mathH \Phi ']\delta_{ij}\\
&+a^{-2}[\nabla^2 (\Psi +\Phi )\delta^i_j-(\Psi + \Phi )_{|^i_j}].
\end{align}\label{eq:perturbed_field_1}
\end{subequations}
The conformal Hubble parameter is ${\mathH = a'/a}$, where prime denotes the derivative with respect to the conformal time $\eta$. In the Eqs.~(\ref{eq:perturbed_field_1}), subscript "$|$" is  the covariant derivative with respect to the spatial 3-dimensional metric and the box operator is ${\Box =\nabla^{\mu}\nabla_{\mu}}$. The linear perturbation of CDM energy momentum tensor becomes $\delta T^{\mu}_{\nu}=\rho_{\rm m} [\Delta_{\rm m} u_{\nu}u^{\mu}+u^{\mu}\delta u_{\nu}+u_{\nu}\delta u^{\mu}]$ and density contrast is represented by $\Delta_{\rm m} \equiv  \delta \rho_{\rm m}/ \rho_{\rm m}$.
By applying Fourier Transformation of the perturbation equations, we find the following perturbed Einstein equations:
\begin{subequations}
\begin{align}
&k^2\tilde{ \Phi} +3\mathH (\tilde{\Phi}' - \mathH \tilde{\Psi})=4\rm{\pi} G_N a^2 \tilde{\Delta}_{\rm m} \rho_{\rm m }\label{eq:pert_fourier_1},\\
&k^2(\tilde{\Phi}' - \mathH \tilde{\Psi})=-4\rm{\pi} G_Na^2\rho_{\rm m}\tilde{\theta},\\
&\tilde{\Psi} =-\tilde{\Phi},\\
&\tilde{\Phi} '' +2\mathH \tilde{\Phi} ' -\mathH \tilde{\Psi}' -(\mathH ^2 +2\mathH ' )\tilde{\Psi} = 0.
\end{align}\label{eq:perturbed_equations_f}
\end{subequations}
The sign $"\tilde{\quad}"$ corresponds to the Fourier mode. The velocity divergence in the Eqs.~(\ref{eq:perturbed_equations_f}) is $\tilde{\theta}\equiv {\bf i} k_{j}\tilde{v}^j$.  Combining Eqs. (\ref{eq:perturbed_equations_f}) and continuity equation, the evolution equation for the CDM density contrast at the linear regime on the \emph{sub-horizon scales} is as follows:
\begin{equation}\label{eq:linear_pert_1}
\frac{\md ^2 \tilde{\Delta}_{\rm m}}{\md N^2}+\left ( \dfrac{\md \ln{\mathH}}{\md N}+1  \right )   \frac{\md \tilde{\Delta}_{\rm m}}{\md N}-\frac{3}{2}\tilde{\Delta}_{\rm m}=0,
\end{equation}
where $N\equiv\ln a$ is the number of $e$-foldings. One can derive a set of differential equations representing the perturbations in the BVDE component. However, we consider suitable initial conditions to suppress the dark energy perturbations and finally, the effect of instability is resolved.

Since  Eq. (\ref{eq:linear_pert_1}) is independent from \emph{scale of structure}, therefore, one can define $\tilde{\Delta}_{\rm m}({\bf{k}},a)\equiv \delta_{\rm m}(a)\tilde{\Delta}_{\rm m}({\bf{k}})$ with  two independent modes which are called \emph{decaying} ($\delta^-_{\rm m}$) and \emph{growing} ($\delta^+_{\rm m}$) modes. The linear growth rate of the density contrast, $f$, which is related to the peculiar velocity in the linear theory is defined by \citep{Peebles1993}:
\bea
f(a)\equiv \frac{\md \ln{D^{+}_{\rm m}(a)}}{\md \ln{a}}
\eea
where $D^+_{\rm m}(a)\equiv \delta^+_{\rm m}(a)/\delta^+_{\rm m}(a=1)$ which is known as \emph{growth function}.  The growth rate measurements are characterized by the peculiar velocities obtained from the Redshift Space Distortion (RSD) observations \citep{Kaiser:1987qv}. To achieve proper observational quantity comparable with the growth rate computed from the linear perturbation theory, one can compare transverse and line of sight anisotropies influenced by the peculiar motion in the redshift space clustering of galaxies. Weak lensing and/or RSD \citep{Song_2009,Nesseris2017} yield a robust combination, namely  $f \sigma_8(z)\equiv f(z) \sigma_8(z)$ which is bias independent observable quantity.  Here the variance of the linear density contrast on scale $R_8 = 8 h^{-1}$ Mpc is $\sigma_8(z)\equiv \sigma (R_8,z)$ and it is given by:
\begin{equation}\label{eq:sigma-rms}
\sigma(R_8, z) = D_{\rm m}^+(z)\left[\int_0^\infty\frac{\md k}{2 \rm{\pi}^2}  k^2 {\mathcal P}_{\rm m}(k) W^2(kR_8) \right]^{1/2}
\end{equation}
where $W(kR) = \frac{3(\sin kR - kR \cos kR)}{(kR)^3} $ and $R=(3M/4 \rm{\pi} \rho_{\rm m})^{1/3}$. The matter power spectrum is introduced by $\langle \tilde{\Delta}_{\rm m}({\bf k}) \tilde{\Delta}_{\rm m}({\bf k}') \rangle=(2\rm{\pi})^3\delta_{D}({\bf k}-{\bf k}')\mathcal{P}_{\rm m}(k)$. Theoretical formula for the present matter power spectrum is ${\mathcal{P}_{\rm m}(k)=\mathcal{P}_0 k^{n_s}T^2(k)}$ where $\mathcal{P}_0$ is normalization constant and $n_s= 0.968\pm 0.006$ according to the recent reanalysis of the \planck data \citep{Ade:2015xua}. $T(k)$ is transfer function. Since, at the very high redshift, our BVDE model is almost similar to that of supposed by \lcdm, therefore, we use the BBKS transfer function model \citep{Bardeen1986}:
\begin{equation} \label{eq:bardeen}
T(k) = C_q \left[ 1+ 3.89 q + (16.1 q)^2 + (5.46 q)^3 + (6.71 q )^4 \right]^{-1/4},
\end{equation}
where $C_q \equiv \ln (1+2.34 q)/2.34 q$ and $q \equiv k/\Gamma$. Here $\Gamma$ is the shape parameter, given by:
\begin{equation}
\Gamma = \Omega^0_{\rm m} \tilde{h} \exp (- \Omega^0_{\rm b} - \sqrt{2 \tilde{h}} \Omega^0_{\rm b}/\Omega^0_{\rm m}).
\end{equation}
It is worth noting that in the general case, one should recalculate the transfer function to find robust results \citep[and references therein]{Wang2010} (e.g. to assess non-linear evolution see \citep{smith2003stable,mcdonald2006dependence,Lewis:2006fu}).
\begin{figure}
\centering
\includegraphics[width=\columnwidth]{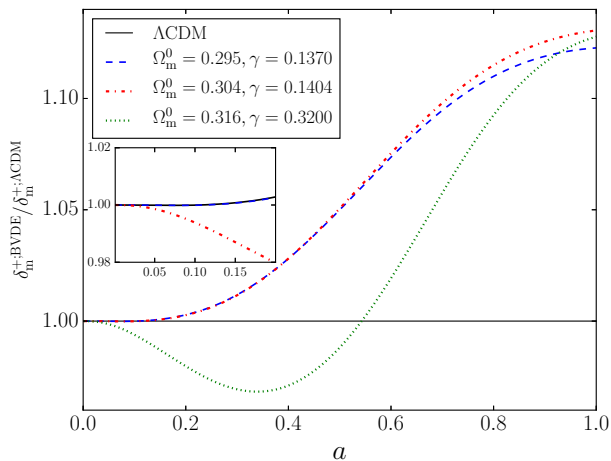}
\caption{The growing mode ratio of BVDE and \lcdm , $\delta_{\rm m}^{+;\text{BVDE}}/\delta_{\rm m}^{+;\Lambda \text{CDM}}$ for different values of free parameters as a function of scale factor. The value of free parameters obtained from the background estimations reported by Ref.~\citet{Mostaghel:2016lcd}. } \label{fig:Normalized_growthrate_ratio}
\end{figure}

\begin{figure}
\centering
\includegraphics[width=\columnwidth]{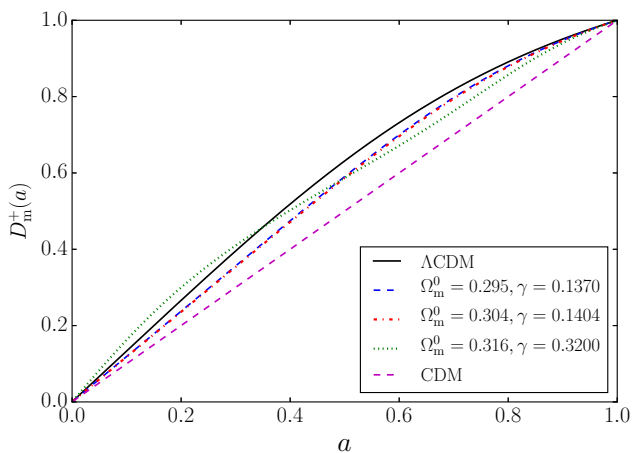}
\caption{ Evolution of CDM growth function as a function of scale factor for various values of free parameters. These values have been set according to background estimations represented in  the Ref.~\citet{Mostaghel:2016lcd}.}\label{fig:CDM_Growing_mode}
\end{figure}

\begin{figure}
\centering
    \includegraphics[width=\columnwidth]{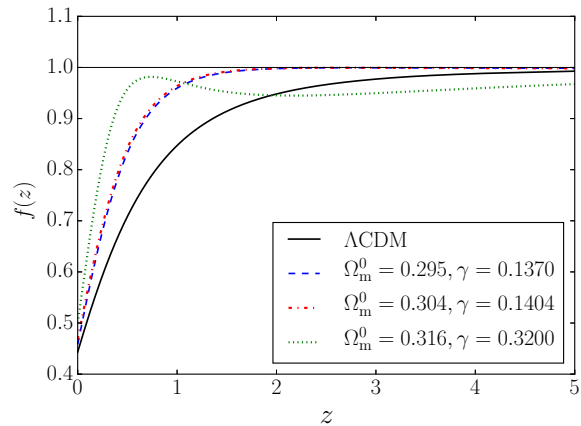}
    \caption{ Growth rate of model for different values of viscosity in comparison of \lcdm.}
\label{fig:growth-rate}
\end{figure}

Fig.~\ref{fig:Normalized_growthrate_ratio} indicates the ratio of BVDE growing mode to the \lcdm model, i.e. $\delta^{+;\rm {BVDE}}_{\rm m}/\delta^{+;\Lambda\text{CDM}}_{\rm m}$. Our result shows that at the early epoch, growing mode in the BVDE model is smaller than that of in \lcdm, on the contrary, at the late time, this quantity becomes higher than growing mode in \lcdm.  However, incorporating bulk viscosity for the dark energy fluid leads to an early dark energy (see Fig. \ref{fig:Om_DE}). Such component for the dark energy with negative pressure, can reduce the rate of large scale structure formation during  the matter dominant Universe.  As illustrated in Fig.~\ref{fig:Normalized_growthrate_ratio}, the growing mode is not monotonic function versus scale factor and  structure formation experiences a delay in the presence of viscous dark energy.

In Fig.~\ref{fig:CDM_Growing_mode}, we plot the evolution of $D^+_{\rm m}(a)$, as a function of scale factor. By increasing the bulk viscosity coefficient in the BVDE model, we find a deviation from the standard \lcdm model. Subsequently, the BVDE cosmological model predicts a different rate of  structure formation, leading to a new observational consequences at relevant redshifts.

 We also depict the linear growth rate in Fig.~\ref{fig:growth-rate}. For almost $\gamma\lesssim\gamma_{\times}\sim 0.36$ 
\citep{Mostaghel:2016lcd}, the early dark energy contribution can not compensate the higher role of cold dark matter at the late time. Consequently, the growth rate is higher than \lcdm.  Such behavior is no longer valid for almost $\gamma\gtrsim\gamma_{\times}$ as illustrated in Fig.~\ref{fig:growth-rate}.  There are more abundance of CDM clusters around $z\sim 0.8$ for higher value of viscosity coefficient. This behavior is due to the slowly varying scale factor while the perturbations have opportunity to grow up \citep{Lahav1991}. We also expect that for higher value of viscosity coefficient, the CMB photons have more chance to experience dynamical gravitational potential at the intermediate scale factor (Fig. \ref{fig:Om_DE}).  In the next section, we will examine the ISW contribution on the CMB map due to the BVDE model, precisely.

\section{Integrated Sachs-Wolfe Effect}\label{sec:ISW}
\begin{figure}
\centering
\includegraphics[width=\columnwidth]{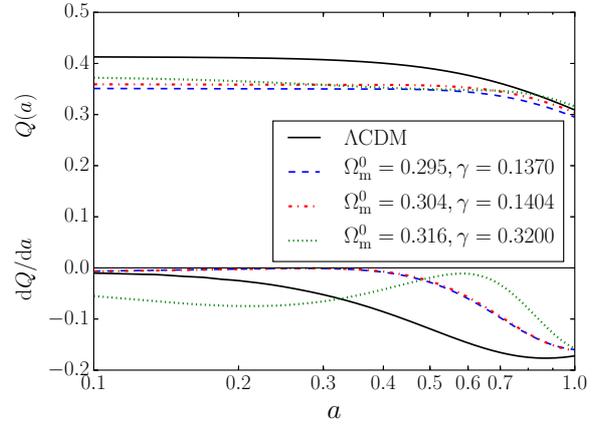}
\caption{{\it Upper part:} The evolution of the $Q(a)$ function for various values of free parameters. {\it Lower part:} The evolution of the $\md Q(a)/\md a$ for the BVDE and \lcdm models. By definition $Q(a=1)=\Omega_{\rm m}^0$. }\label{fig:ISW_1_a}
\end{figure}

One of the sensitive and feasible tracers of the dark energy density is ISW effects. This effect is due to the interaction of CMB photons with time-varying gravitational potential. This term contains all processes due to the non-static metric fluctuations. During radiation and dark energy dominated epochs, the gravitational potential, $\Phi$, has dynamics leading to non-vanishing contribution on the CMB fluctuations. Here, we ignore the non-significant early ISW effect associated with the radiation dominated Universe.  The quantitative formula for computing the ISW contribution on the CMB fluctuation seen in direction $\hat{\bf n}$ reads as \citep{Sachs1967,Seljak:1996is,gordon2004low,afshordi2004cross,ho2008correlation,olivares2008integrated,ade2016planck}:
\begin{eqnarray}\label{ISW1}
\left(\dfrac{\Delta T}{T_{\cmb}}\right)(\hat{\bf n})= -2\int_0^{\chi_{\rm CMB}}\md \chi a^2 H(a) \frac{\partial \Phi(\hat{\bf n},a)}{\partial a},
\end{eqnarray}
here $\chi_{\rm CMB}$ is the comoving distance to the last scattering surface.  Using the Eq.~(\ref{eq:pert_fourier_1}), the Poisson equation in the Fourier space for the scales smaller than the Hubble radius, $k\gg \mathH$, becomes:
\begin{equation}
k^2 \tilde{\Phi} =4\rm{\pi} \GN a^2 \rho_{\rm m}\delta^+_{\rm m}(a)\tilde{\Delta}_{\rm m}({\bf k})\label{eq:ISW_1}
\end{equation}
Finally, we get the gravitational potential as:
\begin{align}
\tilde{\Phi} =\dfrac{3H_0^2}{2k^2}a^2E^2(a)\Omega_{\rm m}(a)\delta^{+}_{\rm m}(a)\tilde{\Delta}_{\rm m}({\bf k}),
\end{align}
where $E^2(a)\equiv H^2(a)/H_0^2$. Now we define the function $Q(a)$ by \citep{Schaefer2008}:
\begin{eqnarray}
Q(a)\equiv a^2 E^2(a)\Omega_{\rm m}(a)D^{+}_{\rm m}(a)\delta^{+}_{\rm m}(a=1),
\end{eqnarray}
Finally, the Eq. (\ref{ISW1}) equates to:
\begin{align}
&\left(\dfrac{\Delta T}{T_{\cmb}}\right)(\hat{\bf n})=\nonumber\\
&-3H_0^2 \int_{0}^{\chi_{\rm CMB}}{\md \chi a^2 H(a) \dfrac{\md Q(a)}{\md a}}\int \frac{\md{\bf k}}{(2\rm{\pi})^{3/2}k^2}\rm{e}^{-i{\bf k}.\hat{\bf n}\chi}\tilde \Delta_{\rm m}(\bf k)
\end{align}
According to this quantitative equation, the $Q(a)$ function and its derivative are the sources of \isw effect \citep{Schaefer2008,Wang2010a,Wang2010}. In our BVDE model, the ISW is altered by modified Hubble parameter and growth function, while the Poisson equation remains unchanged due to the absence of interaction between the BVDE and cold dark components. As indicated in the upper part of  Fig. \ref{fig:ISW_1_a}, the $Q(a)$ function in the BVDE model is lower than the \lcdm model. The $\md Q(a)/\md a$ behaves as non-monotonic function versus scale factor leading to non-trivial contribution in the ISW effect. Actually taking into account the $\gamma$ (dimensionless viscosity coefficient), the magnitude of $\md Q(a)/\md a$ as a source of ISW, becomes smaller than \lcdm. This behavior can be justified by considering the contribution of $\Omega_{\rm DE}$ indicated in Fig. \ref{fig:Om_DE} and also by Fig. \ref{fig:CDM_Growing_mode}. In addition, the behavior of $H(a)$ in BVDE model is different compared to \lcdm  \citep{Mostaghel:2016lcd}. As discussed before, for $\gamma\gtrsim\gamma_{\times}$, we expect to have opposite contribution of BVDE resulting in the higher value for the source of ISW.

\begin{figure}
\centering
\includegraphics[width=\columnwidth]{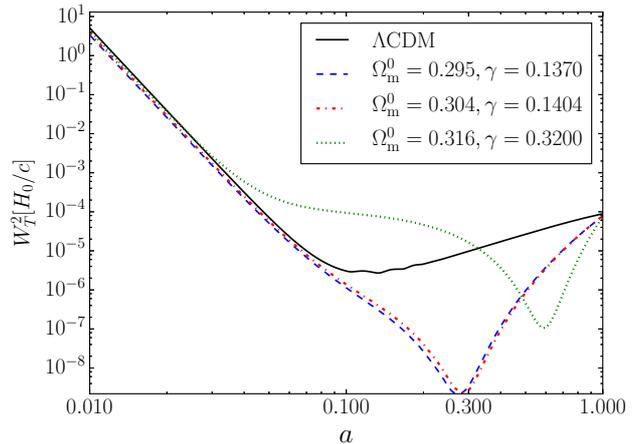}
\caption{Weight function of ISW phenomenon versus scale factor for the BVDE and \lcdm models.}\label{fig:wt}
\end{figure}
Till now we explained the physical model for ISW of CMB fluctuations, but due to stochasticity nature of CMB field, the practical observable measures are inferred from probabilistic framework in the context of n-point auto-correlation and cross-correlation approaches. We turn to the two-point correlation function of CMB temperature fluctuation expanded in terms of spherical harmonic basis functions: $C_T(\theta)=\sum_{\ell}(2\ell +1)\mathcal{C}_{\ell}^{TT} \mathcal{W}_{\ell}^2 P_{\ell}(\cos \theta)/4\rm{\pi}$. Here $\mathcal{C}_{\ell}^{TT}$ is temperature angular power spectrum, $P_{\ell}$ is the Legendre polynomial and $\mathcal{W}_{\ell}$ is a smoothing function\footnote{For a Gaussian kernel function, we have ${\mathcal{W}_{\ell}=\exp\left(-\theta_{\rm beam}^2\ell(\ell+1)/2)\right)}$ and $\theta_{\rm beam}\equiv\theta_{\rm FWHM}/\sqrt{8\ln 2}$.}.
Considering spherical harmonic expansion of $\Delta T/T_{\rm CMB}$ and taking into account Limber projection for simplification\footnote{Actually, the integration over various values of $k$ is replaced by the most dominant contribution term.} \citep{Limber1954,kaiser1992weak}, ISW power spectrum reads as:
\bea\label{eq:iswtt}
\mathcal{C}_{TT,\ell}^{\isw} = \int_0^{\chi_{\rm CMB}} \md \chi\, \mathcal{W}_{\ell}^2H_0^4\frac{W_T^2(\chi) }{\chi^2}\frac{\mathcal{P}_{\rm m}((\ell+1/2)/\chi)}{\left[(\ell+1/2)/\chi\right]^4}       ,
\eea
where $k=(\ell+1/2)/\chi$ \citep{ho2008correlation} and the weighting function including the evolution of gravitational potential is $W_T(\chi)\equiv 3a^2H(a)dQ(a)/da$.  Fig. \ref{fig:wt} indicates $W_T^2$ as a function of scale factor for the BVDE and \lcdm models. For $\gamma\lesssim\gamma_{\times}$, the contribution of bulk viscosity decreases the amount of $W_T^2$ comparing to \lcdm. This is also justified by increasing the contribution of cold dark matter in the BVDE model.  Non-monotonic behavior with respect to viscosity coefficient for  $\gamma\gtrsim\gamma_{\times}$ is a consequence of  previous results shown in Fig. \ref{fig:ISW_1_a}.

\begin{figure}
\centering
\includegraphics[width=\columnwidth]{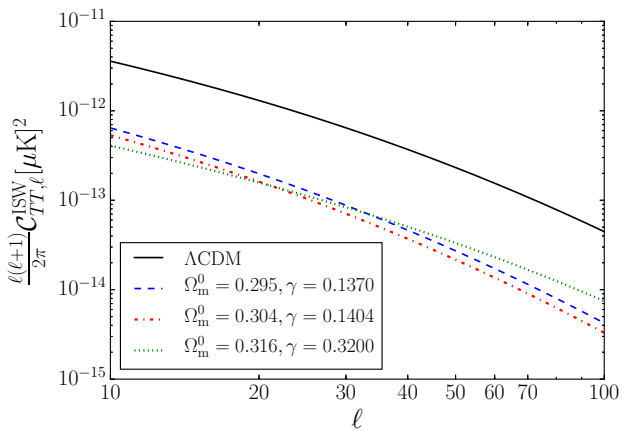}
\caption{
The ISW power spectrum $\mathcal{C}_{TT,\ell}^{\isw}$ versus multiple order $\ell$ for the BVDE and \lcdm models.
}\label{fig:isw-bulk-class-best2}
\end{figure}

Fig.~\ref{fig:isw-bulk-class-best2} illustrates the effect of bulk viscosity on the ISW power spectrum. The ISW power spectrum for BVDE model is lower than \lcdm. Increasing $\gamma$ leads to decrease $\mathcal{C}_{TT,\ell}^{\isw}$ for $\gamma\lesssim\gamma_{\times}$. This behavior is no longer valid for $\gamma\gtrsim\gamma_{\times}$ due to the non-monotonic contribution of viscosity effect in the BVDE model.

Physical interpretation for this behavior is clarified  by looking at the source terms in Eq. (\ref{eq:iswtt}). For $\gamma\lesssim\gamma_{\times}$ the ratio of $\Omega_{\rm m}/\Omega_{\rm DE}$ is higher than that of for \lcdm during long period of evolution. Therefore, the amount of variation in the gravitational potential is less than \lcdm expressed by $\md Q(a)/\md a$ (a representative for $\dot{\Phi}$) in Fig. \ref{fig:ISW_1_a} resulting in the lower value of late ISW.  For $\gamma\gtrsim\gamma_{\times}$, we expect to have more (less) contribution of viscous dark energy at the early (late) time. This manner leads to have a turning point for $\gamma$ dependency of ISW phenomenon in the BVDE model and CMB photons are experiencing more dynamical potential (see Fig. \ref{fig:ISW_1_a}). The reducing in the $\Omega_{\rm DE}/\Omega_{\rm m}$ at the late time and in the $\md Q(a)/\md a$ can not be compensated by the more contribution of the viscous dark energy at the early time leading to have lower ISW with respect to \lcdm.

\begin{figure}
\centering
\includegraphics[width=\columnwidth]{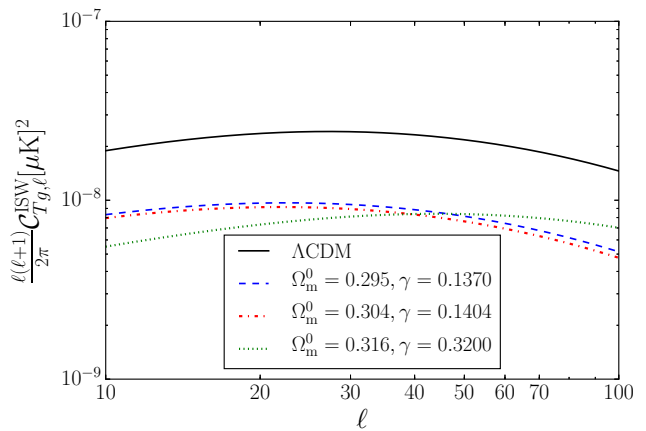}
\includegraphics[width=\columnwidth]{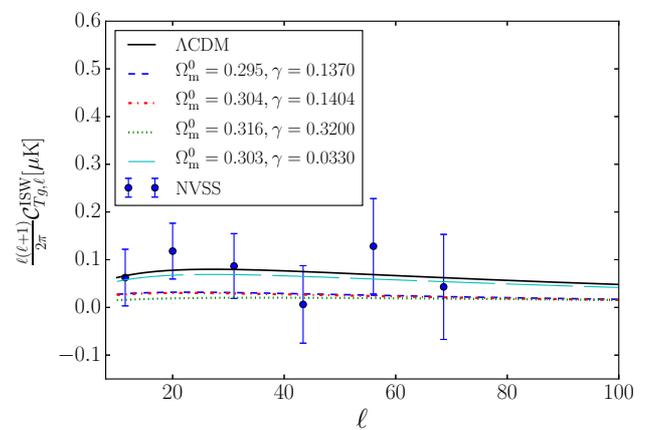}
\caption{ The ISW-cross power spectrum $\mathcal{C}_{Tg,\ell}^{\isw}$ versus multiple order $\ell$ for standard \lcdm and BVDE model. Here other cosmological parameters are fixed. The observed data points from NVSS surveys have been depicted.
}\label{fig:isw-bulk-best-cross}
\end{figure}

The physical processes governing on the CMB fluctuations can be classified into the primary and secondary classes.  Therefore, the observed CMB map is a superposition of mentioned processes having almost different contributions in the various scales with almost same frequency dependent. Practically, it is very hard to distinguish the corresponding footprints. The early ISW effect which is due to the metric perturbation just after photon decoupling epoch on the sub-sound-horizon scales, has mainly contribution on the scale $\ell\sim 200$. On the other hand, late ISW due to evolving gravitational potential opportunity imprints on the small multipoles. Cosmic variance on small $\ell$ as well as the corresponding small value of amplitude are other obstacles for recognizing the ISW phenomenon alone \citep{Wang2010a,ferraro2015wise,Lesgourges2013}.   A way to resolve  this difficulty and constructing an observable power spectrum for ISW, is computing cross-correlation of observed CMB fluctuations with the tracers of the metric perturbations \citep[and references therein]{Wang2010a,Wang2010}. This is because of spatial of ISW signal is correlated with the matter distribution of the local Universe at low redshifts.  To this end, we compute the cross-correlation of $\Delta T/T_{\rm CMB}$ with e.g. density contrast of galaxies or quasars, $\delta_g$.  The scale independent density contrast of observable galaxies is:
\bea\label{eq:galaxy_density_contrast_1}
\delta_g(\hat{\bf n}) = \int{b(z) \frac{\md N}{\md z}\Delta_{\rm m}(\hat{\bf n},z)   \md z  },
\eea
where $b(z)$ is the bias between galaxies and dark matter density perturbations.  Also, $\md N/\md z$ is the selection function of the survey. Similar to the previous strategy, the line of sight integral for density contrast of visible galaxies reads as:
\begin{align}
\delta_g(\hat{\bf n}) &=\int_0 ^{\chi_{\rm CMB}}\md \chi b(a)a^2\frac{\md N}{\md a}H(a)D^+_{\rm m}(a)\delta^+_{\rm m}(a=1)\nonumber\\
 &\quad\times\int\frac{\md{\bf k}}{(2\rm{\pi})^{3/2}}\rm{e}^{-i{\bf k}.\hat{\bf n}\chi}\tilde \Delta_{\rm m}(\bf k)
\end{align}
We define the weight function for observable galaxy density contrast as: ${W_g (\chi)\equiv b(a)a^2(\md N/\md a) H(a)D^+_{\rm m}(a)\delta^+_{\rm m}(a=1)}$.
The angular cross-correlation of CMB fluctuations and visible galaxy density contrast considering only ISW contribution in the flat-sky approximation is given by:
\begin{align}
&\mathcal{C}_{Tg,\ell}^{\isw}=\int_0^{\chi_{\rm CMB}}\md \chi H_0^2 \frac{W_T(\chi)W_g(\chi)}{\chi^2} \frac{\mathcal{P}_{\rm m}((\ell+1/2)/\chi)}{\left[(\ell+1/2)/\chi\right]^2}     ,
\end{align}
In order to compute cross-correlation power spectrum, we use the NRAO VLA Sky Survey (NVSS)\footnote{\texttt{http://www.cv.nrao.edu/nvss/}} sample \citep{Condon1998} which has:
\bea
\left(b(z)\frac{\md N}{\md z}\right)_{\rm NVSS}=b_{\rm eff}\frac{\alpha^{\alpha +1}}{z_*^{\alpha+1}\Gamma(\alpha)}z^{\alpha}\rm{e}^{-\alpha z/z_*}
\eea
where $b_{\rm eff}=1.98$, $z_*=0.79$ and $\alpha=1.18$ \citep{ho2008correlation}. The NVSS catalog includes the North sky of $-40$ deg declination in
1.4 GHz continuum band \citep{Condon1998}. For other observed samples released by different surveys, we should use proper redshift distribution functions as reported  e.g. in \citep{mcdonald2005linear,Giannantonio:2008zi,douspis2008optimising,ho2008correlation,xia2009high,Wang2010a,ferraro2015wise,ade2016planck}.

The \isw cross-power spectrum is show in Fig.~\ref{fig:isw-bulk-best-cross}.  From observational point of view, the value of $\mathcal{C}_{Tg,\ell}^{\isw}$ shows a considerable magnification compared to  $\mathcal{C}_{TT,\ell}^{\isw}$. In the upper panel of Fig.~\ref{fig:isw-bulk-best-cross}, we show the effect of bulk viscosity on the ISW-cross power spectrum. By increasing $\gamma$, the value of the ISW-cross power spectrum goes down and we have a turning point in which the behavior of the ISW-cross power spectrum changes in opposite way but not for all multipoles as revealed by the mentioned figure.  For $\gamma\sim \gamma_{\times}$, at the late time, we have a deficiency in  $\md Q(a)/\md a$ (Fig. \ref{fig:ISW_1_a}), while for almost long interval of scale factor, it gets higher value compared to other cases yielding non-monotonic behavior for ISW cross-power spectrum with galaxies. This behavior provides an opportunity to get more consistent theoretical predictions with relevant observations. The lower panel of Fig.~\ref{fig:isw-bulk-best-cross} illustrates the $\mathcal{C}_{Tg,\ell}^{\isw}$ in $\mu$K scale compared to the NVSS \citep{ho2008correlation}.

In the next section, we will examine the observational consistency of BVDE model based on observational quantities derived in the perturbations approached.

\section {Observational consistency}\label{sec:observation}

In our previous paper, we mainly concentrated on the background evolution and derived the best fit values for model free parameters. To make more complete present discussion on the contribution of bulk viscous dark energy model in the dark matter perturbations evolution, we focus on an observable quantity coming from perturbation formalism, namely $f\sigma_8(z)$. Table \ref{table:fsigma8} contains the observed value of $f\sigma_8(z)$ with associated $1\sigma$ uncertainty according to the "Gold-2017" catalogue \citep[see references therein]{Nesseris2017}. Our relevant model free parameters are $\gamma$ and $\Omega_{\rm m}^0$ with constant priors. The Hubble parameter at present time is mainly constrained by SNIa+BAO+{\it Planck} reported by  \cite{Mostaghel:2016lcd}.

\begin{table}
\centering
\begin{tabular}{|c|c|c|}
\hline
Index & Redshift & $f\sigma_{8,{\rm obs}}$ \\
\hline\hline
1&$0.02$ & $0.428\pm0.0465$\\
2&$0.02$ & $0.398\pm0.065$\\
3&$0.02$ &$0.314\pm 0.048$\\
4&$0.10$ & $0.370\pm 0.130$\\
5&$0.15$ & $0.490\pm 0.145$\\
6&$0.17$ & $0.510\pm 0.060$\\
7&$0.18$ & $0.360 \pm 0.090$\\
8&$0.25$ & $0.3512\pm 0.0583$\\
9&$0.32$ & $0.384\pm 0.095$\\
10&$0.37$ & $0.4602\pm 0.0378$\\
11&$0.38$ & $0.440\pm 0.060$\\
12&$0.44$ &$0.413\pm 0.080$\\
13&$0.59$ & $0.488 \pm 0.060$\\
14&$0.60$ &$0.550\pm 0.120$\\
15&$0.60$ &$0.390\pm 0.063$\\
16&$0.73$ & $0.437\pm 0.072$\\
17&$0.86$ & $0.400\pm 0.110$\\
18&$1.40$ & $0.482\pm 0.116$\\
\hline
\end{tabular}
\caption{The current observational value of the $f\sigma_8(z)$ according to the "Gold-2017" catalogue \citep{Nesseris2017}.}
\label{table:fsigma8}
\end{table}

\begin{figure}
\centering
\includegraphics[width=\columnwidth]{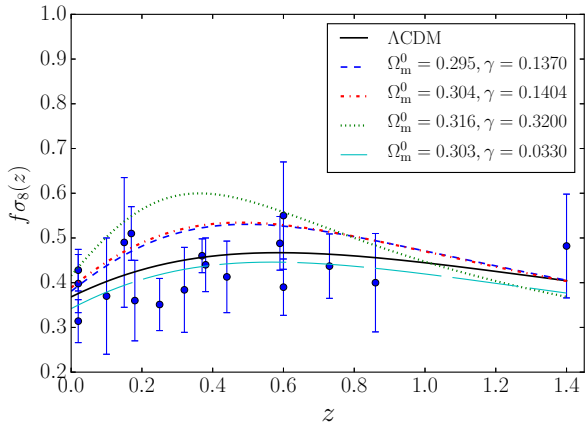}
\caption{\label{fig:fsigma8} The $f\sigma_8$ as a function of redshift for various values of  BVDE model free parameters. The solid line corresponds to the best fit parameters based on \lcdm. Increasing the viscous coefficient ($\gamma$) shows ups and downs in the behavior of  $f\sigma_8$ leading to have more consistency with observational data points.    }
\end{figure}

\begin{figure*}
\centering
\includegraphics[width=\textwidth]{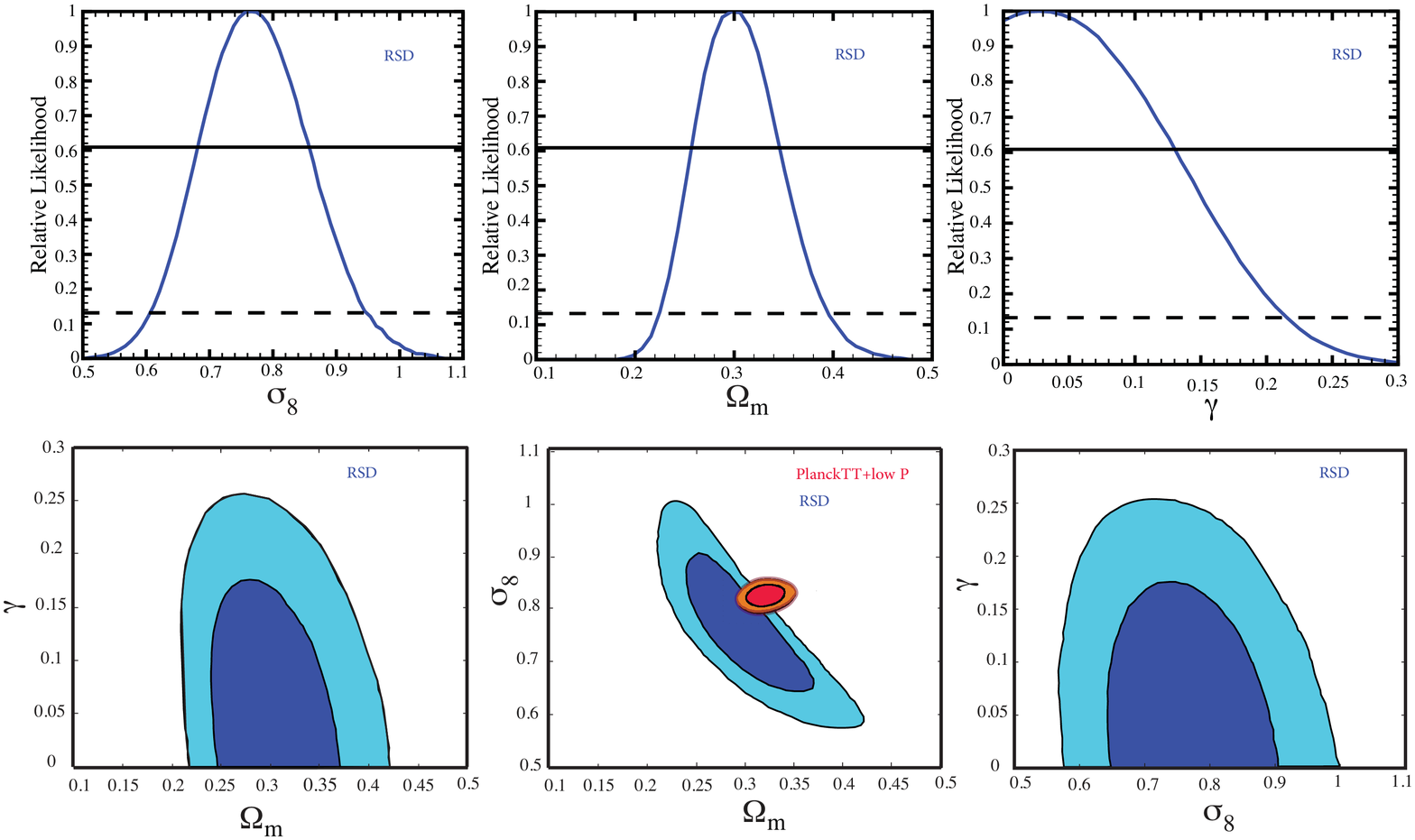}
\caption{\label{fig:all}, Upper panels corresponds to marginalized posterior function for BVDE model free parameters. Marginalized confidence regions at $68\%$ and $95\%$ confidence levels. The small contours is devoted to \lcdm model. }
\end{figure*}

In order to compare the observational data set with that of predicted by our model, we utilize likelihood function with the following $\chi^2$:
\begin{align}
\chi^2_{\rm RSD} \equiv \Delta f\sigma_8 ^{\texttt t}\cdot {C}^{-1}\cdot \Delta f\sigma_8
\end{align}
where $\Delta f\sigma_8\equiv f\sigma_{8}^{\rm obs}(z)-f\sigma_{8}^{\rm the}(z;\Omega^0_{\rm m},\gamma) $ and ${C}$ is the covariance matrix of RSD data set. The best fit values for BVDE free parameters based on RSD observation are reported in  Table \ref{tb:rsd}.  Fig. \ref{fig:fsigma8} indicates the behavior of $f\sigma_8$ as a function of redshift for various values of model free parameters. The symbols correspond to the most new catalog including observational values.  The marginalized likelihood function for $\sigma_8$, $\Omega^0_{\rm m}$ and $\gamma$ determined by RSD observations are depicted in the upper panel of Fig. \ref{fig:all}. The lower panel of Fig. \ref{fig:all} indicates  the contour plots illustrating the marginalized confidence regions at $68\%$ and $95\%$ levels. Taking into account the bulk viscosity for dark energy component, manipulates the growing of the dark matter in the Universe decreasing tension in the present fluctuation spectrum, $\sigma_8$, which has been mentioned in \citep{Ade:2015xua,Heymans:2012gg,erben2013cfhtlens}.

\begin{table}
\centering
\begin{tabular}{|c|c|}
\hline
\hline Parameter & RSD  \\
\hline $\Omega^0_{\rm m}$ &$0.303^{+0.044+0.093}_{-0.038-0.070}$\\
\hline $\gamma$ &$0.033^{+0.098+0.182}_{-0.033-0.033}$   \\
\hline $\sigma_8$ & $0.769^{+0.080+0.181}_{-0.089-0.154}$ \\ \hline
\end{tabular}
\caption{Best fit values for BVDE model using RSD data at $68 \%$  and $95\%$ confidence intervals.}\label{tb:rsd}
\end{table}

\section{Summary and conclusion}\label{sec:summary}
In this paper, following our previous paper on proposing a new dynamical dark energy model inspired by thermodynamical dissipative phenomena, we examined the linear perturbation theory of the dark matter in the presence of viscous dark energy model. In order to probe the dark energy properties, the clustering of large scale structure and ISW have been elucidated.

Taking into account viscosity coefficient for our BVDE model, suppresses the $\delta^+_{\rm m}$ comparing to the same quantity computed for \lcdm at early epoch. While dark matter growing mode is boosted at the late time. For higher value of the viscous coefficient, we obtained higher value of scale factor for which the $\delta^+_{\rm m}$ becomes higher than  \lcdm (see Fig. \ref{fig:Normalized_growthrate_ratio}). 
 As illustrated in Fig.~\ref{fig:Normalized_growthrate_ratio}, the dark matter growing mode is not monotonic function versus scale factor and  structure formation experiences a delay in the presence of viscous dark energy.

Fig. \ref{fig:CDM_Growing_mode} indicated the value of growth function ($D^+_{\rm m}$). By increasing the bulk viscosity coefficient in the BVDE model, we found a deviation from the standard \lcdm model. Therefore we expect to have a manipulation in the $\sigma_8$ introduced by Eq.  (\ref{eq:sigma-rms}).  The linear growth rate in Fig.~\ref{fig:growth-rate} indicated that for almost $\gamma\lesssim\gamma_{\times}$ \citep{Mostaghel:2016lcd}, the early dark energy contribution can not compensate the higher role of cold dark matter at the late time. Consequently, the growth rate is higher than \lcdm.  Such behavior is no longer valid for almost $\gamma\gtrsim\gamma_{\times}$.  We found a bump in $f(z)$ around $z\sim 0.8$ for higher $\gamma$.

ISW power spectrum for our BVDE model has been illustrated in Fig.~\ref{fig:isw-bulk-class-best2} confirming non-monotonic contribution of viscosity effect in the BVDE model. To give a physical interpretation for this behavior, we should look at the contribution of $\Omega_{\rm DE}$, $H$, $D^+_{\rm m}(a)$, $Q(a)$ and $\md Q(a)/\md a$. The bulk viscosity parameter can extensively modify the rate of Hubble expansion and consequently, it mainly affects the evolution of the dark matter density parameter and $D^{+}_{\rm m}(a)$. In another word, incorporating bulk viscosity for our dark energy fluid, reduces $\Omega_{\rm DE}/\Omega_{\rm m}$ for long period of cosmic evolution. According to Fig. \ref{fig:Normalized_growthrate_ratio}, the growing mode of matter is higher than \lcdm. The source term in ISW is diminished considerably leading to have more static gravitational potential and therefore less ISW. This behavior is not monotonic with respect to $\gamma$. For almost $\gamma\gtrsim\gamma_{\times}$, at intermediate range of scale factor, the BVDE model effectively reduces the structure formation giving higher ISW term. It is worth mentioning that, such magnification can never compensate the sharp reduction in $W_T^2$ and finally we get the lower ISW value comparing to the \lcdm model (Figs. \ref{fig:ISW_1_a}, \ref{fig:wt} and \ref{fig:isw-bulk-class-best2}). Indeed, incorporating viscosity for $\gamma\lesssim \gamma_{\times}$ damps the growing mode of the newtonian potential slower than \lcdm model.

To magnify the amount of ISW signal in one hand and on the other hand to reduce the degeneracies in determining the source of fluctuations in the power spectrum, we computed the ISW cross-correlation with visible galaxy density contrast.  Such cross-correlation can magnify  the value of $\mathcal{C}_{Tg,\ell}^{\isw}$ compared to  $\mathcal{C}_{TT,\ell}^{\isw}$. In BVDE model, the value of the ISW-cross power spectrum decreases by increasing $\gamma$ and we found a turning point confirming non-monotonic behavior of ISW-cross power spectrum.  Fig.~\ref{fig:isw-bulk-best-cross} illustrates the $\mathcal{C}_{Tg,\ell}^{\isw}$ in $\mu$K scale compared to the NVSS.

To examine the observational consistency, we have utilized "Gold-2017" RSD observations. The value of $f\sigma_8$ as a function of redshift has been indicated in Fig. \ref{fig:fsigma8}.
Our posterior analysis indicated that $\Omega^0_{\rm m} =0.303^{+0.044+0.093}_{-0.038-0.070}$, $\gamma=0.033^{+0.098+0.182}_{-0.033-0.033}$ and $\sigma_8= 0.769^{+0.080+0.181}_{-0.089-0.154}$ at $1\sigma$ and $2\sigma$ level of confidences, respectively. It seems that our model could reduce the tension in $\sigma_8$ (see Fig. \ref{fig:all}).

Finally, considering the contribution of coupling between dark sectors of the Universe incorporating viscosity for the dark energy could be interesting and utilizing structure formation with background observations would be able to distinguish between such families of models \citep{Schaefer2008}. Taking into account such interaction may provide an opportunity to examine the stability of underlying viscous dark energy model. To resolve  acausal problem for the bulk viscous dark energy fluid, one can keep the collision time scale in the transport equation. We will address them in the future work.
\section*{acknowledgements}
We are really grateful to Bj\"{o}rn Malte Sch\"{a}fer for his very constructive comments.  We also thank to Nima Khosravi and Marzieh Farhang for their suggestions on the framework of this paper. SMSM is grateful to school of Physics, Institute for research in fundamental sciences (IPM), where some parts of this paper have been finalized. 



\bibliography{bib}{}
\bibliographystyle{mnras}

\bsp	
\label{lastpage}
\end{document}